\documentclass[12pt,a4paper]{article}
\usepackage{a4wide}
\usepackage{epsfig}
\usepackage{cite}
\usepackage{amsmath}
\usepackage{graphicx}
\usepackage{latexsym}
\usepackage{array}

\newlength{\absize}
\def\gsim{\mathrel{\rlap{\raise 2.5pt \hbox{$>$}}\lower 2.5pt
\hbox{$\sim$}}}
\def\lsim{\mathrel{\rlap{\raise 2.5pt \hbox{$<$}}\lower 2.5pt
\hbox{$\sim$}}}

\usepackage[dvips]{color}
\definecolor{Black}{named}{Black}
\definecolor{Red}{named}{Red}

\begin{document}

\thispagestyle{empty}
\renewcommand{\thefootnote}{\fnsymbol{footnote}}
\newpage\normalsize
\pagestyle{plain}
\setlength{\baselineskip}{4ex}\par
\setcounter{footnote}{0}
\renewcommand{\thefootnote}{\arabic{footnote}}
\newcommand{\preprint}[1]{%
\begin{flushright}
\setlength{\baselineskip}{3ex} #1
\end{flushright}}
\renewcommand{\title}[1]{%
\begin{center}
\LARGE #1
\end{center}\par}
\renewcommand{\author}[1]{%
\vspace{2ex}
{\Large
\begin{center}
\setlength{\baselineskip}{3ex} #1 \par
\end{center}}}
\renewcommand{\thanks}[1]{\footnote{#1}}
\renewcommand{\abstract}[1]{%
\vspace{2ex}
\normalsize
\begin{center}
\centerline{\bf Abstract}\par
\vspace{2ex}
\parbox{\absize}{#1\setlength{\baselineskip}{2.5ex}\par}
\end{center}}

\vspace*{4mm} %\vfill

\title{Spin identification of heavy nonstandard
bosons in dilepton and diphoton events at the LHC} \vfill

\author{P. Osland,$^{a,}$\footnote{E-mail: per.osland@ift.uib.no}
A. A. Pankov,$^{b,}$\footnote{E-mail: pankov@ictp.it} 
N. Paver$^{c,}$\footnote{E-mail: nello.paver@ts.infn.it} and 
A. V. Tsytrinov$^{b,}$\footnote{E-mail: tsytrin@rambler.ru}}

\begin{center}
$^{a}$Department of Physics and Technology, University of Bergen,
Postboks 7803, N-5020  Bergen, Norway\\
$^{b}$The Abdus Salam ICTP Affiliated Centre, Technical
University of Gomel, 246746 Gomel, Belarus\\
$^{c}$University of Trieste and INFN-Trieste Section, 34100
Trieste, Italy
\end{center}
%
%\keywords{Polarized $e^+e^-$ collisions, Models beyond the standard model}
%
%\pacs{12.60.-i, 12.60.Rc, 12.60.Cn}
%
\vfill

\abstract{
New Physics scenarios generally predict the existence of
very heavy quantum states that can possibly manifest
themselves as peaks in the cross sections at the LHC.
For values of the parameters in certain domains,
different nonstandard models can generate peaks with
the same mass and same number of events. In this case,
the spin determination of a peak, requiring the angular
analysis of the events, becomes crucial in order to
identify the relevant nonstandard source.
We here discuss, using a particularly suitable
symmetrically integrated angular asymmetry applied to
Drell-Yan dilepton and diphoton events at LHC, the
identification reach on the exchanges in these reactions
of the following heavy bosons: spin-2 Randall-Sundrum
graviton excitations; spin-1 heavy neutral gauge
bosons $Z^\prime$; and spin-0 SUSY $R$-parity violating
sneutrinos.}

\section{Introduction}
The occurrence of the heavy bosons predicted by
models beyond the standard model (SM), with mass
scales $M\gg M_{W,Z}$, can be signalled by the
observation of (narrow) peaks in the cross sections for
reactions among standard model particles at the high
energies available at the LHC. However, the
observation of a peak/resonance at some large mass
$M=M_R$ may not be sufficient to identify its underlying
nonstandard model, in the multitude of potential
sources of such a signal. Indeed, in ``confusion regions''
of the parameters, different models can give the same
$M_R$ and same number of events under the peak. In
that case, the test of the peak/resonance quantum
numbers, in the first place of the spin, is needed to
discriminate the models against each other in the
confusion regions. Specifically, one defines
for the individual nonstandard scenarios a
{\it discovery reach} as the maximum value of
$M_R$ for peak observation over the SM 
background, and an {\it identification reach}
as the maximum value of $M_R$ for which the model can be
unambiguously discriminated from the other competing
ones as the source of the peak.

Particularly clean signals of heavy neutral
resonances are expected in the inclusive reactions
at the LHC:
\begin{equation}
p+p\to l^+l^-+X\ \ (l=e,\mu)\ \ \ {\rm and}
\ \ \ p+p\to\gamma\gamma+X,
\label{proc}
\end{equation}
where they can show up as peaks in the dilepton (and diphoton)
invariant mass $M$. While the total resonant cross section
determines the number of events, hence the discovery reaches on
the considered models, the angular analysis of the events allows
to discriminate the spin-hypotheses from each other, due to the
(very) different characteristic angular distributions. In the next
sections we discuss the identification of the spin-2, spin-1 and
spin-0 hypotheses, modelled by the Randall-Sundrum model with one
warped extra dimension \cite{Randall:1999ee}, a set of $Z^\prime$
models \cite{Langacker:2008yv}, and the $R$-parity violating
sneutrino exchange \cite{Kalinowski:1997bc}, respectively.

\section{Cross sections and center-edge asymmetry}
The total cross section for a heavy resonance
discovery in the events (\ref{proc}) at an invariant
dilepton (or diphoton) mass $M=M_R$ (with
$R=G,Z^\prime, {\tilde\nu}$ denoting
graviton, $Z^\prime$ and sneutrino, respectively) is:
\begin{equation}
\sigma{(pp\to R)} \cdot {\rm BR}(R \to l^+l^-)
=\int_{-z_{\rm{cut}}}^{z_{\rm cut}}{\rm d} z
\int_{M_{R}-\Delta M/2}^{M_{R}+\Delta M/2}{\rm d} M
\int_{y_{\rm min}}^{y_{\rm max}}{\rm d} y
\frac{{\rm d}\sigma}{{\rm d} M\, {\rm d} y\, {\rm d} z}.
\label{TotCr}
\end{equation}
Resonance spin-diagnosis makes use of the comparison
between the different differential angular
distributions \cite{Allanach:2000nr,Cousins:2005pq}:
\begin{equation}
\frac{{\rm d}\sigma}{{\rm d} z} =
\int_{M_{R}-\Delta M/2}^{M_{R}+\Delta M/2}{\rm d} M
\int_{y_{\rm min}}^{y_{\rm max}}
\frac{{\rm d}\sigma}{{\rm d} M\, {\rm d} y\, {\rm d} z}\,{\rm d} y.
\label{DiffCr}
\end{equation}
In Eqs.~(\ref{TotCr}) and (\ref{DiffCr}),
$z=\cos\theta_{\rm cm}$ and $y$ are the lepton-quark
(or photon-quark) angle in the
dilepton (or diphoton) center-of-mass frame and the dilepton
rapidity, respectively, and cuts on phase
space due to detector acceptance are indicated.
Furthermore, $\Delta M$ is an invariant mass bin
around $M_R$, reflecting the detector energy resolution,
see for instance Ref.~\cite{Atlas}. To evaluate the number
$N_S$ of resonant signal events, time-integrated
luminosities of 100 and 10 ${\rm fb}^{-1}$ will
be assumed, as well as 90\% reconstruction 
efficiencies for both electrons and muons and 80\% for
photons \cite{Cousins:2004jc}.
Typical experimental cuts are:
$p_\bot >20$~GeV and pseudorapidity
$\vert\eta\vert < 2.5$ for both leptons;
$p_\bot >40$~GeV and
$\vert\eta\vert < 2.4$ for photons.
To evaluate Eqs.~(\ref{TotCr}) and (\ref{DiffCr}), 
the parton subprocess cross sections will be
convoluted with the CTEQ6 parton distributions of
Ref.~\cite{Pumplin:2002vw}. Next-to-leading QCD
effects can be accounted for by $K$-factors,
and for simplicity of the presentation we here adopt a
flat value $K=1.3$ for all considered processes.

In practice, due to the completely symmetric $pp$
initial state, the event-by-event determination
of the sign of $z$ may at the LHC be not fully
unambiguous. This difficulty may be avoided
by using as the basic observable for the angular
analysis the $z$-evenly integrated center-edge
angular asymmetry, defined as
\cite{Dvergsnes:2004tw,Osland:2008sy,Osland:2009tn,Osland:2010yg}:
\begin{equation}
\label{ace}
A_{\rm{CE}}=\frac{\sigma_{\rm{CE}}}{\sigma}\quad{\rm
with} \quad \sigma_{\rm{CE}} \equiv \left[\int_{-z^*}^{z^*} -
\left(\int_{-z_{\rm cut}}^{-z^*} +\int_{z^*}^{z_{\rm
cut}}\right)\right] \frac{{\rm d} \sigma}{{\rm d} z}\, {\rm d} z.
\end{equation}
In Eq.~(\ref{ace}), $0<z^*<z_{\rm cut}$ defines
the separation between the ``center'' and the
``edge'' angular regions and is {\it a priori}
arbitrary, but the numerical analysis shows that
it can be ``optimized'' to $z^*\simeq 0.5$. The
additional advantage of using $A_{\rm CE}$ is that,
as being a ratio of integrated cross
sections, it should be much less sensitive to
systematic uncertainties than the ``absolute''
distributions (examples are the $K$-factor
uncertainties from different possible sets of
parton distributions and from the choice of
factorization vs renormalization mass scales).

\section{Nonstandard interactions and relevant angular
distributions}
We list, for the nonstandard models of
interest here, the basic features relevant to the
angular analysis and the spin-identification.

\subsection{RS model with one compactified
extra dimension}
The simplest version \cite{Randall:1999ee},
originally proposed as a rationale for the
gauge hierarchy problem
$M_{\rm EW}\ll M_{\rm Pl}$, consists of one
warped extra spatial coordinate $y$ with exponential
warp factor $\exp{(-k\pi\vert y\vert)}$ (with $k>0$
the 5D curvature assumed of order $M_{\rm Pl}$), and
two three-dimensional branes placed at a compactification
distance $R_c$ in $y$. The SM fields are localized to
the so-called TeV brane, while gravity originates
on the other one, the so-called Planck brane,
but is allowed to propagate in the
full 5D space. The consequence of the chosen
space-time geometry is that, in the reduction
to four dimensions, a Planck-brane mass spectrum
with characteristic scale of order
${\bar M}_{\rm Pl}=1/\sqrt{8\pi G_{\rm N}}\simeq2.4\times 10^{15}$ GeV,
is exponentially ``warped'' down to the TeV-brane, and
the cut-off on the effective theory becomes there
$\Lambda_\pi={\bar M}_{\rm Pl}\exp{(-k\pi R_c)}$.
For $k R_c\simeq 12$, $\Lambda_\pi$ is of the TeV order
and this opens up the appealing possibility of observing
gravitational effects at the LHC energies.
Notably, these signatures consist of a tower 
of spin-2 graviton excitations that can be exchanged 
in processes (\ref{proc}) and show up as narrow peaks 
in $M$ with the specific mass spectrum
$M_n=x_nk\exp{(-k\pi R_c)}$, of order
$\Lambda_\pi\sim\, {\rm TeV}$
($x_n$ are the roots of $J_1(x_n)=0$), and
couplings to SM particles of order $1/\Lambda_\pi$.

The model can be conveniently parametrized in terms
of $M_G$, the mass of the lowest graviton excitation,
and of the ``universal'' dimensionless coupling
$c=k/{\bar M}_{\rm Pl}$. Theoretically, the
expected ``natural'' ranges are
$0.01< c < 0.1$ and $\Lambda_\pi< 10$ TeV
\cite{Davoudiasl:1999jd}. Current 95\% CL
experimental limits \cite{Abazov:2010xh} are in the range
$M_G>600$ GeV (for $c\cong 0.01$) up to
$M_G>1.05$ TeV (for $c\cong 0.1$).

For dilepton production, in self-explaining
notations and with $\epsilon^G_q$, $\epsilon^G_g$
and $\epsilon_q^{\rm SM}$ the fractions
of $G$-events under the $M_R$-peak initiated
by $q{\bar q}$, $gg$ and the SM background,
respectively, the $z$-even distributions needed 
in (\ref{ace}) can at the leading order be
expressed, as \cite{Han:1998sg}: %Giudice:1998ck}:
\begin{equation}
\frac{{\rm d}\sigma^G}{{\rm d}z}=
\frac{3}{8}(1+z^2)\sigma_{q}^{\rm SM} +
\frac{5}{8}(1-3z^2+4z^4)\sigma^G_{q} +
\frac{5}{8}(1-z^4)\sigma^G_{g},
\label{Diffg}
\end{equation}
and:
\begin{equation}%\label{ACE2}
A_{\rm CE}^{G} =\epsilon_q^{\rm SM}\,A_{\rm CE}^{\rm SM} +
\epsilon^G_q\left[2\,{z^*}^5+\frac{5}{2}\,z^*(1-{z^*}^2)-1\right]
+ \epsilon^G_g\left[\frac{1}{2}\,{z^*}(5-{z^*}^4)-1\right].
\label{aceg}
\end{equation}
For the diphoton events, the leading
order RS resonance exchange contributions
to $q {\bar q}\to G \to \gamma\gamma$ and
$gg \to G \to \gamma\gamma$ can analogously be written
as \cite{Sridhar:2001sf}:
\begin{equation}
\frac{{\rm d}\sigma^G}{{\rm d}z} =
\frac{5}{8} (1-z^4) \sigma_q^G +
\frac{5}{32} (1 + 6 z^{2} + z^{4}) \sigma_g^G,
\label{Diffggam}
\end{equation}
and
\begin{equation}
\label{aceggam}
A_{\rm CE}^{G} =
\epsilon_q^G \left[\frac{1}{2} z^{*} (5-{z^{*}}^{4}) - 1\right] +
\epsilon_g^G \left[-1 + \frac{5}{8}z^{*} + \frac{5}{4}{z^{*}}^{3} +
\frac{1}{8}{z^{*}}^{5}\right].
\end{equation}
Next-to-leading order QCD effects, and the
corresponding $K$-factors, have been evaluated in
Ref.~\cite{Mathews} and in Ref.~\cite{Kumar} for the
dilepton and the diphoton channels, respectively. One
important remark is that for the diphoton channel,
due to spin-1$\not\to\gamma\gamma$, the viable
hypotheses reduce to spin-2 and spin-0 exchanges only and,
moreover, the RS model makes the definite prediction
${\rm BR}(G\to\gamma\gamma)/{\rm BR}(G\to l^+l^-)\simeq 2$
\cite{Han:1998sg}.

\subsection{Heavy neutral gauge bosons}
The spin-1 hypothesis is in process (\ref{proc}) realised by
$q{\bar q}$ annihilation into lepton pairs through $Z^\prime$
intermediate states \cite{Langacker:2008yv}. Such bosons are
generally predicted by electroweak models beyond the SM, based on
extended gauge symmetries. Generally, $Z^\prime$ models depend on
$M_{Z^\prime}$ and on the left- and right-handed couplings to SM
fermions. In the sequel, results will be given for a popular class
of models for which the values of these couplings are fixed
theoretically, so that only $M_{Z^\prime}$ is a free parameter.
These are the $Z^\prime_\chi$, $Z^\prime_\psi$, $Z^\prime_\eta$,
$Z^\prime_{\rm LR}$, $Z^\prime_{\rm ALR}$ models, and the
``sequential'' $Z^\prime_{\rm SSM}$ model with $Z^\prime$
couplings identical to the $Z$ ones. Current experimental lower
limits (95\% CL) on $M_{Z^\prime}$ depend on models, and range
from 878 GeV for $Z^\prime_\psi$ up to 1.03 TeV for $Z^\prime_{\rm
SSM}$ \cite{Aaltonen:2008ah}.

The $z$-even angular distributions for the 
partonic subprocesses 
$q{\bar q}\to {Z^\prime}\to l^+l^-$ have the
same form as in the SM and, therefore, the
resulting $A_{\rm CE}$ is the same for 
all $Z^\prime$ models:
\begin{equation}
\frac{{\rm d}\sigma^{Z^\prime}}{{\rm d}z}=\frac{3}{8}
(1+z^2)[\sigma_{q}^{\rm SM}+ \sigma^{Z^\prime}_{q}];
\label{Diffzprime}
\end{equation}

\begin{equation}
\label{acezprime}
A_{\rm CE}^{Z^\prime}\equiv A_{\rm CE}^{\rm SM}=
\frac{1}{2}z^*(z^{*2}+3)-1.
\end{equation}
Consequently, the $A_{\rm CE}$-based angular
analysis should have a considerable degree
of $Z^\prime$ model independence.
A discussion of next-to-leading QCD corrections
can be found, for instance, in
Ref.~\cite{Carena:2004xs}.

\subsection{$R$-parity violating sneutrino exchange}
$R$-parity is defined as $R_p=(-1)^{(2S+3B+L)}$, and
distinguishes particles from their superpartners.
In scenarios where this symmetry can be violated,
supersymmetric particles can be singly produced from
ordinary matter. In the dilepton process
(\ref{proc}) of interest here, a spin-0 sneutrino
can be exchanged through the subprocess
$d{\bar d} \to{\tilde\nu}\to l^+l^-$ and
manifest itself as a peak at $M=M_{\tilde\nu}$ with
a flat angular distribution \cite{Kalinowski:1997bc}:
\begin{equation}
\frac{{\rm d}\sigma^{\tilde\nu}}{{\rm d} z}=
\frac{3}{8}(1+z^2)\sigma^{\rm SM}_{q}+ \frac{1}{2}\sigma
^{\tilde\nu}_{q},
\label{Diffsneu}
\end{equation}
\begin{equation}
A_{\rm CE}^{\tilde\nu} = \epsilon_q^{\rm SM}\,A_{\rm CE}^{\rm SM}
+\epsilon^{\tilde\nu}_{q}(2z^*-1).
\label{acesneu}
\end{equation}
Results on higher QCD orders and 
supersymmetric QCD corrections available
in the literature indicate the possibility of
somewhat large $K$-factors 
\cite{Choudhury:2002aua,Dreiner:2006sv}.
The cross section is
proportional to the $R$-parity violating product
$X=(\lambda^\prime)^2B_l$ where $B_l$ is the
sneutrino leptonic branching ratio
and $\lambda^\prime$ the relevant sneutrino
coupling to the $d{\bar d}$ quarks. Current
limits on the relevant $\lambda^\prime$s are
of the order of $10^{-2}$ \cite{Barbier:2004ez},
and the experimental 95\% CL lower limits
on $M_{\tilde\nu}$ range from 397 GeV
(for $X=10^{-4}$) to 866 GeV
(for $X=10^{-2}$) \cite{Aaltonen:2008ah}.
We may take for $X$, presently not really
constrained for sneutrino masses of order
1 TeV or higher, the (rather generous) interval
$10^{-5}<X<10^{-1}$.

\section{Spin-diagnosis using $A_{\rm CE}$}

\begin{figure}[h]
\begin{center}
\includegraphics[width=6.5in]{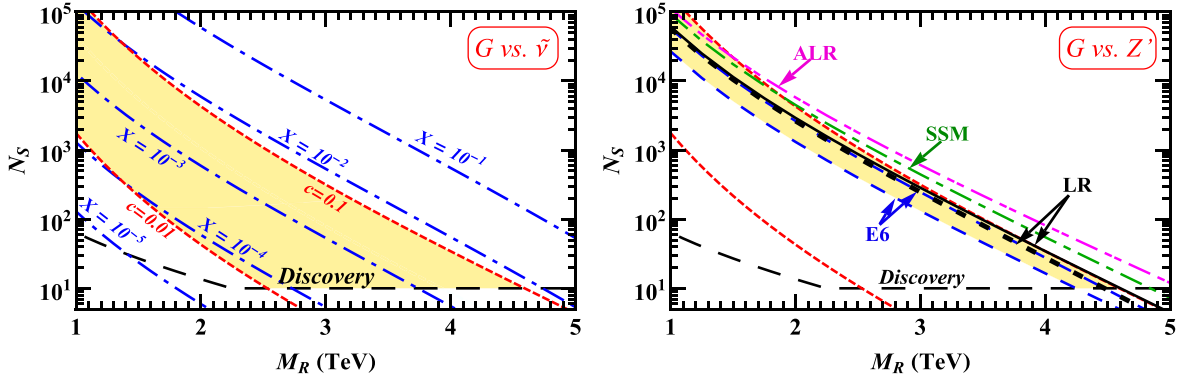}%\hspace{2pc}%
\hspace*{-10mm}
\end{center}
%\begin{minipage}[b]{14pc}
\caption{\label{fig1}Discovery and confusion regions (yellow) vs $M_R$.}
\hspace*{-10mm}
%\end{minipage}
\end{figure}

The nonstandard models briefly described in the previous section
can mimic each other as sources of an observed peak in $M$, for
values of the parameters included in so-called ``confusion
regions'' (of course included in their respective experimental
and/or theoretical discovery domains), where they can give the same
number of signal events $N_S$. The $M_R$-$N_S$ plots in
Fig.~\ref{fig1} show as examples the graviton vs sneutrino and
graviton vs $Z^\prime$ confusion domains as well as the number of
events needed for 5-$\sigma$ discovery at the 14 TeV LHC with
luminosity ${\cal L}_{\rm int}=100\, {\rm fb}^{-1}$.
\hspace*{-10mm}

In such confusion regions, one can try to
discriminate models from from one another by means of
the angular distributions of the events, directly
reflecting the different spins of the exchanged particles.
We continue with the examples of confusion
regions in Fig.~\ref{fig1} and start from the
assumption that an observed peak at $M=M_R$ is
the lightest spin-2 graviton (thus, $M_R=M_G$).
We define a ``distance'' among models accordingly:
\begin{equation}
\Delta A_{\rm CE}^{Z^\prime}=A_{\rm CE}^G-A_{\rm CE}^{Z^\prime}
\qquad {\rm and} \qquad
\Delta A_{\rm CE}^{\tilde\nu}=
A_{\rm CE}^G-A_{\rm CE}^{\tilde\nu}.
\label{deltaGSV}
\end{equation}
To assess the domain in the ($M_G,c$) plane
where the competitor spin-1 and spin-0 models
giving the same $N_S$ under the peak
can be {\it excluded} by the starting
RS graviton hypothesis, a simple-minded
$\chi^2$-like criterion can be applied, which
compares the deviations (\ref{deltaGSV}) with the
statistical uncertainty $\delta A_{\rm CE}^G$
pertinent to the RS model (systematic
uncertainties can easily be included). We
impose the two conditions
\begin{equation}
\chi^2\equiv
\vert\Delta A_{\rm CE}^{{Z^\prime},{\tilde\nu}}/ \delta A_{\rm CE}^G \vert^2
> \chi^2_{\rm CL}.
\label{chisquare}
\end{equation}
Here, $\chi^2_{\rm CL}$ specifies a desired exclusion 
confidence level (3.84 for 95\% CL). This condition 
determines the minimum number of events, $N_S^{\rm min}$, 
needed to exclude the spin-1 and spin-0 hypotheses 
(hence to establish the graviton spin-2), and this in
turn will determine the RS graviton {\it identification} 
domain in the ($M_G,c$) plane. Of course, an analogous 
procedure can be applied to the identification of 
$Z^\prime$ and $\tilde\nu$ exchanges against the two 
competing ones. 
In the next section we review the results obtained for 
the three spin-identification analyses based on $A_{\rm CE}$. 
Recents attempts based on, alternatively, $y$-integrated 
asymmetries have been proposed in Ref.~\cite{Diener:2009ee}.   
\begin{figure}[h]
\begin{center}
\includegraphics[width=6.5in]{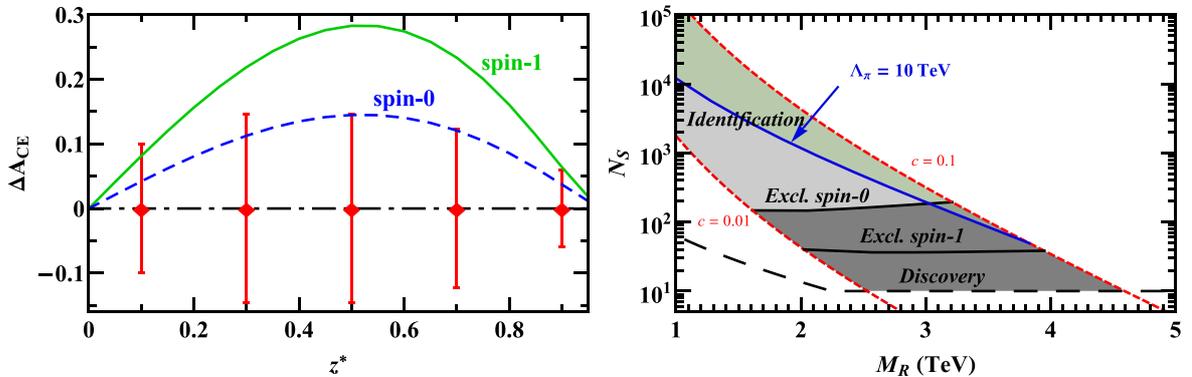}\hspace{2pc}%
\end{center}
%\begin{minipage}[b]{14pc}
\caption{\label{fig2}Deviations (\ref{deltaGSV}) vs $z^*$ (left panel); 
$N_S^{\rm min}$ for spin-1 and spin-0 exclusion from RS graviton 
hypothesis (right panel).}
%\end{minipage}
\end{figure}

\subsection{Spin-2 identification}
Figure~\ref{fig2} shows, for LHC 
energy and luminosity the same as for Fig.~\ref{fig1}, 
the deviations (\ref{deltaGSV}) vs $z^*$ 
for $M_G=1.6$ TeV and 
$c=0.01$, assuming the same $M_R$ and number of peak events 
for the spin-1 $Z^\prime$ and the spin-0 $\tilde\nu$ 
hypotheses. The error bars are the statistical 2-$\sigma$
uncertainties on $A_{\rm CE}^{G}$. 
Figure~\ref{fig2} shows, as anticipated, that $z^*\simeq 0.5$ is
``optimal'', in the sense that at this value there is maximal
sensitivity to the deviations among models and, moreover, the
$\chi^2$ is found to be rather smooth. By imposing 
the conditions (\ref{chisquare}), one
finds the minimum number of events $N_S^{\rm min}$ 
vs $M_G$ (and with $0.01<c<0.1$), needed to exclude 
at 95\% CL the spin-1 as well as the spin-0 hypotheses 
once the spin-2 one has been assumed to be ``true''. 
Such $N_S^{\rm min}$ are reported in Fig.~\ref{fig2}, 
right panel. Notice from this figure that the 
``theoretically favored'' region is severely restricted
to the domain within the $\Lambda_\pi=10\, {\rm TeV}$ 
and the $c=0.1$ contours (green).

\begin{figure}[h]
\begin{center}
\includegraphics[width=5.5in]{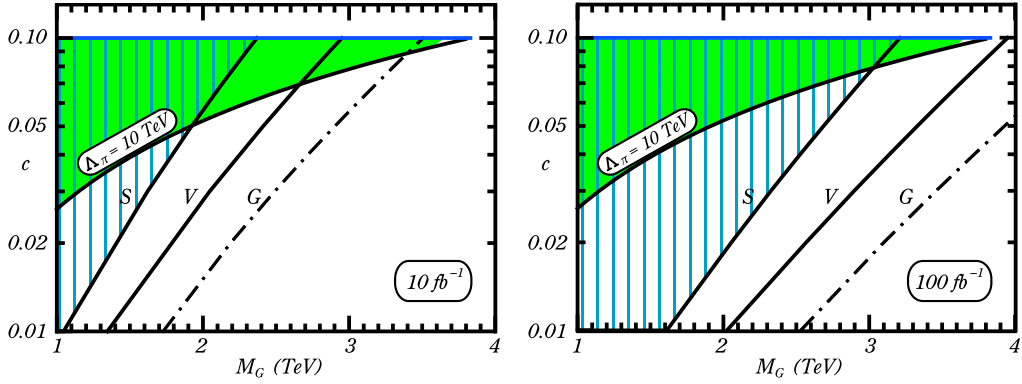}%\hspace{2pc}%
\end{center}
%\begin{minipage}[b]{14pc}
\caption{\label{fig3}RS graviton discovery and identification from dilepton events.}
%\end{minipage}
\end{figure}

\begin{figure}[h]
\begin{center}
\includegraphics[width=5.5in]{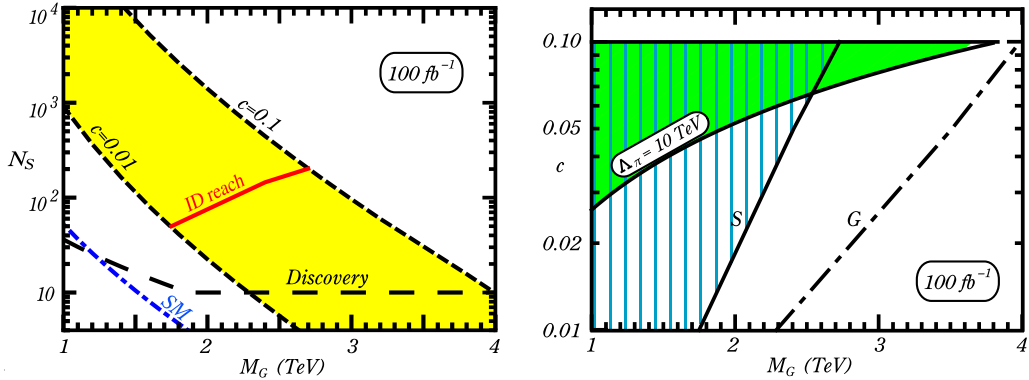}%\hspace{2pc}%
\end{center}
%\begin{minipage}[b]{14pc}
\caption{\label{fig4} RS graviton discovery and identification from diphoton events.}
%\end{minipage}
\end{figure}

Figure~\ref{fig3} shows the expected lowest lying
graviton identification domain at 95\% CL in the
$(M_G,c)$ plane from dilepton events
($l=e,\mu$ combined) at 14 TeV with time-integrated 
luminosities of 10 and 100 
${\rm fb}^{-1}$ \cite{Osland:2008sy}.
Basically, in this figure, the domain to the
left of the line ``$G$'' is the discovery domain;
that to the left of the ``$V$'' line is the
exclusion domain of the $Z^\prime$ hypothesis;
and that to the left of the ``$S$'' line represents
the domain where the $\tilde\nu$ hypothesis
(as well as the $Z^\prime$) can be excluded,
hence the spin-2 identified. From the two panels of
Fig.~\ref{fig3} one can read the expected graviton
identification limits: $M_G<1.1$ or 2.4 TeV for $c=0.01$
or 0.1, respectively, at 10 ${\rm fb}^{-1}$;
$M_G<1.6$ or 3.2 TeV for $c=0.01$ or 0.1,
respectively, at 100 ${\rm fb}^{-1}$. The identification
reach could therefore be a significant portion of the
discovery domain, especially for the higher luminosity. On
the other hand, the discovery domain is really constrained
by the condition $\Lambda_\pi<10$ TeV, if applied
literally.

Figure~\ref{fig4} shows a preliminary attempt to assess
the 95\% CL identification reach on the RS spin-2 graviton
excitation from the diphoton events in (\ref{proc}), by 
means of the $A_{\rm CE}$ analysis, for
${\cal L}_{\rm int}=100$ ${\rm fb}^{-1}$ and cuts and
photon reconstruction efficiencies as outlined in the
Introduction. In this case, only a hypothetical spin-0
resonance decaying to two photons must be excluded.
The curves in those plots must be interpreted
analogously to those in Fig.~\ref{fig3}. Specifically,
the left panel shows the $N_S^{\rm min}$ vs $M_G$ for RS 
identification (or scalar hypothesis rejection) within
$0.01<c<0.1$, while the right panel shows the
identification domain in the $(M_G,c)$ plane. This
tentative example shows that diphoton events might have an 
identification sensitivity to the RS graviton
comparable to the dilepton ones, with the
spin-1 automatically excluded.

\subsection{Spin-1 $Z^\prime$ identification}
Due to our choice of a family of models where the values of the
$Z^\prime$ coupling constants to quarks and leptons have
theoretically pre-determined values, in the ($M_R-N_S$) plot in
Fig.~\ref{fig1} these scenarios are simply represented by lines,
with now $M_R=M_{Z^\prime}$. The figure shows that, at the LHC
luminosity assumed there, some models, namely, the 
$Z^\prime_{\rm ALR}$ and the $Z^\prime_{\rm SSM}$ can be 
discriminated from the RS spin-2 resonance (but not from 
the spin-0 $\tilde\nu$) already
at the level of event rates. The other $Z^\prime$s share 
confusion regions with both spin-2 and spin-0 hypotheses.

The $A_{\rm CE}$-based angular analysis can
be applied quite similar to the preceding case,
this time assuming that an observed peak in
$M$ is due to a $Z^\prime$, and evaluating the
minimal number of events needed for excluding the
spin-2 and spin-0 hypotheses. At the 100
${\rm fb}^{-1}$ luminosity assumed in
Fig.~\ref{fig1}, $N_S^{\rm min}$ turns out to
be about 130 and 200 for exclusion of spin-2 and
spin-0, respectively. This information can easily
be turned into identification limits in terms
of the relevant $M_{Z^\prime}$. For the 14 TeV 
LHC nominal energy and luminosity 100 
${\rm fb}^{-1}$, one could establish the $Z^\prime$ 
hypothesis (by exclusion of spin-2 and spin-0) for 
$M_{Z^\prime}\leq 3.0-3.8$ TeV, depending on 
the particular model. In addition one can make 
pairwise comparisons (hence obtain identification) 
between the considered $Z^\prime$ models with 
same $Z^\prime$ mass on the 
basis of the different expected statistics, in the 
1--2 TeV range for $M_{Z^\prime}$. Details are discussed 
in Ref.~\cite{Osland:2009tn}.

\subsection{Spin-0 sneutrino identification}
Figure~\ref{fig1} shows that the domain in 
the $R$-breaking parameter $X$ 
allowed to sneutrinos is so 
large that its 
discovery domain fully includes those of 
the RS resonance (with $0.01<c<0.1$) and 
of all $Z^\prime$s. The situation would be 
exactly the same even if we restricted 
$X$ to the narrower interval $10^{-4}$--$10^{-2}$. 
\begin{figure}[h]
\begin{center}
\includegraphics[width=5.5in]{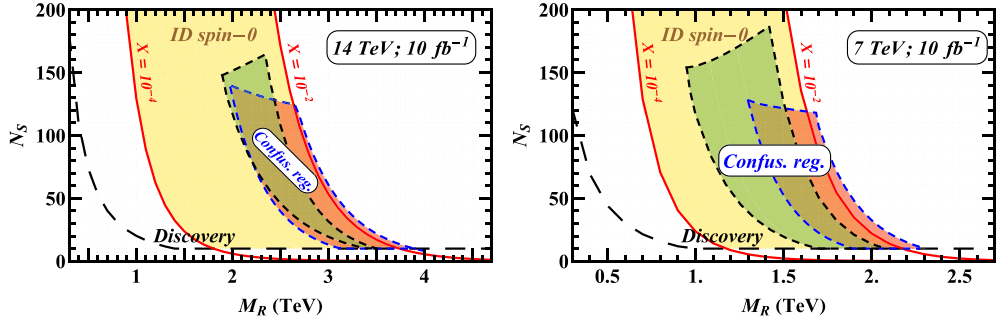}%\hspace{2pc}%
\end{center}
%\begin{minipage}[b]{14pc}
\caption{\label{fig5} Sneutrino discovery and identification regions.}
%\end{minipage}
\end{figure}

Figure~\ref{fig5} shows, as an example, the sneutrino 
confusion regions with RS and $Z^\prime$s vs 
$M_{\tilde\nu}$ for 10 ${\rm fb}^{-1}$, with  
LHC energy 14 TeV (left panel) and 7 TeV (right 
panel), respectively. The $Z^\prime$ models 
are not all explicitly represented, the relevant 
curves lie in the domain between the rightmost 
($Z^\prime_{ARL}$) and the leftmost 
$Z^\prime_\psi$ dashed ones. The condition 
$\Lambda_\pi<10$ TeV is not reported here. 
One can easily read off the minimal number of 
events vs $M_{\tilde\nu}$ needed 
for 95\% CL exclusion of the RS resonance, of the 
spin-1 $Z^\prime$ hypotheses, and both, once 
a peak in dilepton events has been attributed to 
sneutrino exchange in (\ref{proc}). One finds 
that $N_S^{\rm min}\simeq 150$ events are needed 
for sneutrino identification via $A_{\rm CE}$, 
the relevant values of $M_{\tilde\nu}$ being 
constrained to the ranges 1.9--2.7 TeV and 
1.1--1.7 TeV for LHC energies 14 TeV and 7 TeV, 
respectively. At 14 TeV and the highest luminosity 
${\cal L}_{\rm int}=100$ ${\rm fb}^{-1}$, the 
range in $M_{\tilde\nu}$ would be 3.0--3.8 TeV.
A larger number of events would be necessary 
if the condition $\Lambda_\pi<10$ TeV on the RS 
model were applied literally. A more delailed 
numerical analysis is reported in 
Ref.~\cite{Osland:2010yg}.

\end{document}